\documentclass[%
reprint,
superscriptaddress,
floatfix,
amsmath,
amssymb,
aps,
notitlepage,
nofootinbib
]{revtex4-1}
\usepackage{mathrsfs}
\usepackage{graphicx,setspace}
\usepackage{bm}
\usepackage[hidelinks=true]{hyperref}
\usepackage{xcolor}
\definecolor{greenish}{rgb}{0.13,0.58,0.16}
\definecolor{reddish}{RGB}{174,12,48}
\definecolor{blueish}{rgb}{0.12, 0.56, 1.0}
\definecolor{magenta}{rgb}{0.8, 0.0, 0.8}
\hypersetup{
colorlinks   = true, 
urlcolor     = greenish, 
linkcolor    = blueish, 
citecolor   = magenta 
}

\def\grad{\nabla}

\def\D{\mathcal{D}}

\def\u{\mathbf{u}}

\def\phat{\hat{\mathbf{p}}}

\def\ch{\hat{c}}

\def\rhoh{\hat{\varrho}}

\def\x{\mathbf{x}}

\def\R{\mathsf{R}}

\def\st{\text{ss}}
\def\max{\text{max}}

\def\epsilon{\chi}
\def\D{\mathsf{D}}
\def\V{\mathsf{V}}

\begin{document}
	\title{\textit{Rheomergy}: Collective behavior mediated by active flow-based recruitment}
	\author{S Ganga Prasath}
	\affiliation{School of Engineering and Applied Sciences, Harvard University, Cambridge MA 02138.}
	\author{L Mahadevan}
	\email{lmahadev@g.harvard.edu}
	\affiliation{School of Engineering and Applied Sciences, Harvard University, Cambridge MA 02138.}
	\affiliation{Department of Physics, Harvard University, Cambridge MA 02138.}
	\affiliation{Department of Organismic and Evolutionary Biology, Harvard University, Cambridge 02138}

	\date{}

	\begin{abstract}
    The physics of signal propagation in a collection of organisms that communicate with each other  both enables and limits how active excitations at the individual level reach, recruit and lead to collective patterning. Inspired by the patterns in a planar swarm of bees that release pheromones, and use fanning flows to recruit additional bees, we develop a theoretical framework for patterning via active flow-based recruitment. Our model generalizes the well-known Patlak-Keller-Segel model of diffusion-dominated aggregation and leads to more complex phase space of patterns spanned by  two dimensionless parameters that measure the scaled stimulus/activity and the scaled chemotactic response. Together these determine the efficacy of signal communication that leads to a variety of migration and aggregation  patterns consistent with observations.
	\end{abstract}

	\maketitle

    In groups of individual agents in both natural and artificial settings, communication is central to achieving a coordinated collective response. Indeed, the survival of the colony pivots around robust communication in the context of foraging~\cite{couvillon2012dance, jackson2006communication, holldobler1978ethological}, protection against invasion by an intruder~\cite{nouvian2016defensive, trhlin2011chemical}, brood care in eusocial clusters~\cite{gordon1996organization, holldobler1990ants} and other tasks inside the nest~\cite{greene2003cuticular, free1987pheromones}. The efficacy of communication is limited by the dynamics of signal propagation among agents through the environment~\cite{wingreen2006cooperation,seeley2009wisdom, gernat2018automated,prasath2021ant}. Over the last century, stigmergy has emerged as one of the most well studied mechanisms to achieve robust communication~\cite{theraulaz1999brief, perna2017social,camazine2003self}: here animals modify the environment by using localized signals that serve as recruitment cues in the context of functional behavior. However, stigmergy is just one way of environmentally modulated communication. Organisms use slow diffusive signals ~\cite{patlak1953random,keller1970initiation,hillen2009user} that underlie the principle of quorum sensing~\cite{waters2005quorum}, passive and active fluid flows~\cite{ocko2015feedback,jost2007interplay,mathijssen2019collective}, and even elastic deformations~\cite{peleg2018collective} to transmit information via physical communication channels. Indeed, social insects such as honey bees (e.g. the species \textit{apis mellifera}) use all these different modes of communication and actively modify their local mechanical, thermal and hydrodynamic environment to transmit information and coordinate their activities and functions.

    \begin{figure}
    \centering
    \includegraphics[width=0.48\textwidth]{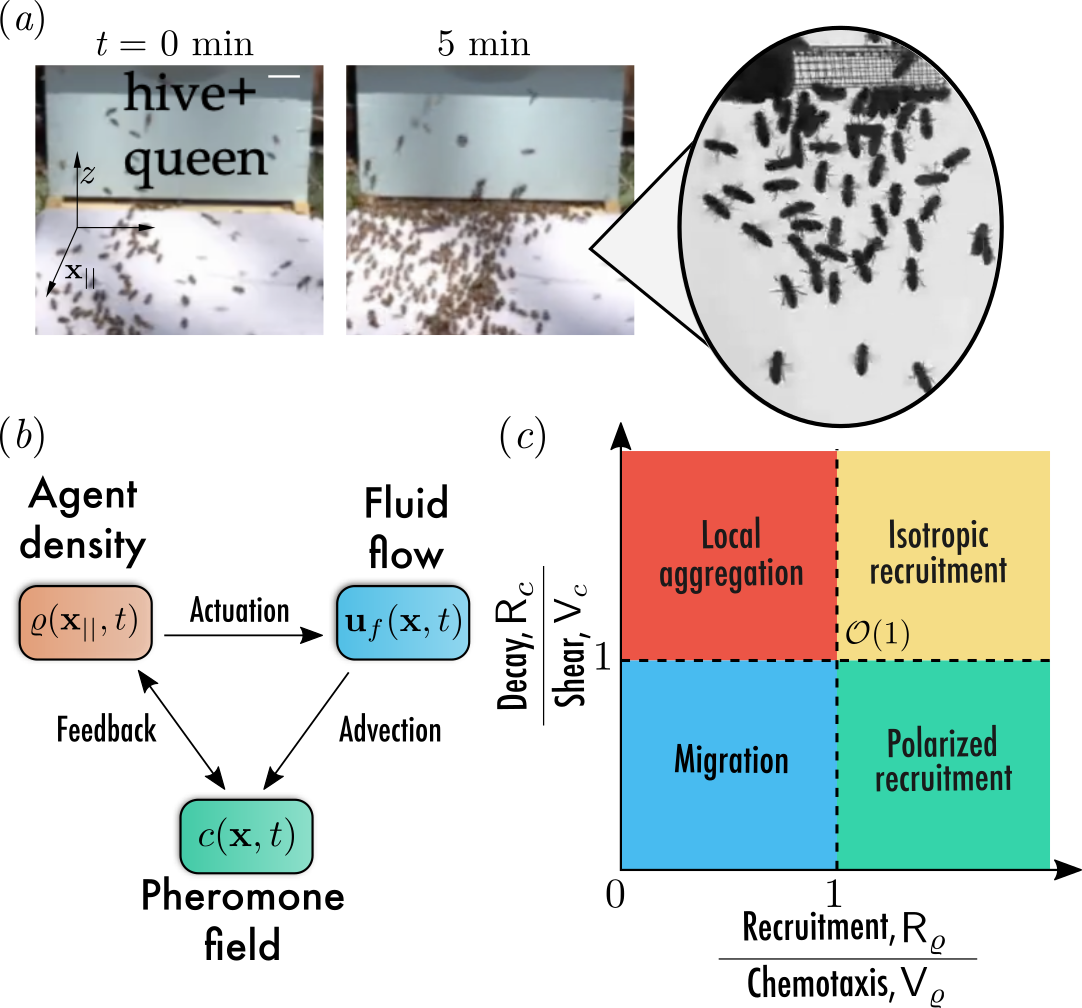}
    \caption{$(a)$ Sequence of images showing recruitment of worker bees by collective fanning to the nest where the queen resides (Image: Peleg Lab at UC Boulder). We see a zoomed in view close to the queen's cage (see also \cite{nguyen2021flow} and ~\cite{youtube}). $(b)$ Interaction rules between the variables in our model in Eqs.~\ref{eq:rhob}-\ref{eq:incomp}: agent density $\varrho(\x_{||}, t)$ which is confined to a 2D surface $\x_{||}= (x,y)$ generates a pheromone field of concentration $c(\x, t)$ that gets fanned through the environment by the actively generated bee velocity field $\u_f(\x, t)$. The advected pheromone causes either additional agents (bees) to be recruitment or causes them to migrate. $(c)$ Phase diagram of \textit{rheomergy} represented using two non-dimensional ratios: Active Peclet number which is the ratio of pheromone decay rate to active fluid shear rate, $\R_c/\V_c$ and recruitment strength to chemotaxis strength, $\R_\varrho/\V_\varrho$ (see text for details). Four phases characterizing the collective behavior of agents exist depending on the region in phase-space: polarized recruitment, migration, local aggregation and isotropic recruitment (similar to that seen in the Patlak-Keller-Segel model).}
    \label{fig:phasediag}
    \end{figure}
    A particularly striking example of active, long-range and fast communication is seen in bees that settle on a substrate and flap their wings to generate a fluid flow parallel to their body axis while releasing pheromones into it~\cite{peters2019collective,nguyen2021flow,reinhard2009role,peters2017wings, peters2017wings,nguyen2021flow,youtube}. In Fig.~\ref{fig:phasediag}$(a)$ and SI Fig.~S1, we show how fluid flow mediated communication leads to  a characteristic triangular swarm shape that migrates towards the source of pheromones~\cite{nguyen2021flow}. These patterns arise on time scales of minutes spontaneously from a disordered state and can extend to include hundreds of bees. 
    
    To understand the nature and efficacy of flow-mediated patterning, we need to couple the three observable spatiotemporal fields that are shown schematically in Fig.~\ref{fig:phasediag}$(b)$: the bee density $\varrho(\x_{||},t)$ (and orientation $\phat(\x_{||},t)$ denotes head forward) as a function of location on the plane $z=0$ denoted by $\mathbf{x}_{||}= (x,y)$ and time, the scalar pheromone concentration field $c(\x,t)$ and the active fluid flow generated by the bees $\u_f (\x, t)$. Bees disperse pheromones from their Nasonov glands into the active flow that they generate by fanning. This leads to the active recruitment of bees from the environment and their migration which then leads to an enhanced rate of pheromone dispersal. The positive feedback leads to spatio-temporal patterns of growing, self-focusing and migrating swarms. While there are many qualitative observations of the process~\cite{reinhard2009role,winston1991biology}, the quantitative collective dynamics of recruitment, agent-environment feedback, and inter-agent interaction mediated by volatile pheromones is largely unexplored.	Here, we develop a minimal model for fluid flow mediated communication and recruitment which we dub \textit{rheomergy}. As we will see,  our model broadens the scope of the classical Patlak-Keller-Segel (PKS) model~\cite{patlak1953random,keller1970initiation,hillen2009user} for chemotaxis by accounting for dynamics of active fluid flow and the behavioral response of the agents to flow-mediated communication leading to a more complex phase space of patterns.
    
    \noindent \textit{Continuum model of rheomergy:} The evolution equations for the agent density $\varrho(\x_{||},t)$, the pheromone field $c(\x,t)$ and the fluid flow generated by the agents, $\u_f (\x, t)$ can be written as,
    \begin{align}
    	\partial_t \varrho + \nabla_{||} \cdot (\u_a \varrho) =& \
    	D_a \nabla_{||}^2 \varrho + (1 - \varrho/\varrho_\max) [k_b \delta \varrho \nonumber \\
    	& \ + \alpha_p \Theta (\delta c) ],\label{eq:rhob} \\
    	\u_a(\x_{||},t)=&\ \epsilon \grad_{||} c, \label{eq:chemov}\\
    	\partial_t c + \grad \cdot(\u_f c) =&\  D_p \nabla^2 c + k_+ \varrho - k_- c, \label{eq:cp}\\
    	\partial_t \u_f =& \ -\frac{\grad p}{\rho_f} + \nu \nabla^2 \u_f - \nu_s \u_f, \label{eq:uf} \\
    	\nabla \cdot \u_f =& \ 0. \label{eq:incomp}
    \end{align}
    Equation~\ref{eq:rhob} describes the migration, relaxation and recruitment of bees as a consequence of the agents relaxing to a nominal ambient density; here the chemotactic velocity defined by Eq.~\ref{eq:chemov} is $\u_a(\x_{||},t)$ with a chemotactic gain $\chi$ and $\grad_{||}$ denotes in-plane gradients, defined along the plane $z=0$, the first term on the right side captures diffusion of agents with a diffusivity $D_a$, the second term denotes relaxation to a nominal density $\varrho_o$ with $\delta \varrho = (\varrho_o - \varrho)$ and $k_b$ denotes the relaxation rate, and the last term characterizes recruitment with $\alpha_p$ being the sensitivity to pheromone above a threshold $c^*$ where $\delta c = (c-c^*)$, and $\Theta(\bullet)$ is the Heaviside function. The agent density is limited to a maximum $\varrho_\max$ through the term $(1 - \varrho/\varrho_\max)$. Equation~\ref{eq:cp} characterizes the dynamics of the pheromone that is advected by the active flow velocity $\u_f$ generated by the swarm, diffuses into the environment with a diffusivity $D_p$, is produced by bees at a rate $k_+$ and decays at a rate  $k_-$. Equations~\ref{eq:uf}-\ref{eq:incomp} follow from the dynamics of (incompressible) flow of fluid with viscosity $\nu$ and density $\rho_f$, and an effective {friction factor} $\nu_s$ associated with fluid motion relative to the substrate. Since this flow is actively generated by bees, we model this using a shear rate at the boundary proportional to their local density: $\partial_z \u_f(z=0) = \beta \varrho \phat $, where $\phat$ is the local polarity of the bees which generates a flow velocity opposite to their orientation and parallel to the surface (see Fig.~\ref{fig:phasediag}$(a)$ and~\cite{peters2017wings}), with $\beta>0$. Equations~\ref{eq:rhob}-\ref{eq:incomp} form a closed system  for the variables $\{ \varrho, c, \u_f, p \}$ which are complete when we specify initial and boundary conditions for all the fields (see SI sec.~III for details).
    
    We make Eqs.~\ref{eq:rhob}-\ref{eq:incomp} dimensionless using the natural length-scale $\zeta^{-1}=(\nu/\nu_s)^{1/2}$ and time-scale $k_b^{-1}$ for Eq.~\ref{eq:rhob}, and $k_-^{-1}$ for Eq.~\ref{eq:cp} (see SI tab.~I for other length and time-scales in the system), the homogeneous relaxation density $\varrho_o$ as the density-scale, the maximum pheromone concentration $c_o$ as the concentration-scale (see SI sec.~I for details), resulting in 7 non-dimensional numbers associated with scaled magnitudes of chemotactic strength, $\V_\varrho =(\chi c_o \zeta^2)/k_b$; agent diffusivity, $\D_\varrho=(D_a \zeta^2/k_b)$; recruitment strength, $\R_\varrho = \alpha_p c_o/(k_b \varrho_o)$; shear, $\V_c = (\beta \varrho_o/k_-)$; pheromone decay, $\R_c = k_+ \varrho_o/(k_- c_o)$; pheromone diffusivity, $\D_c = (D_p \zeta^2/k_-)$; friction factor, $\D_u = \nu_s/(\beta \varrho_o)$.
    This allows us to rewrite the variables after scaling as: $\tilde{\varrho} \rightarrow \varrho/\varrho_o, \tilde{c} \rightarrow c/c_o, \tilde{\u}_f \rightarrow \u_f (\zeta/ \beta \varrho_o), \tilde{p} \rightarrow p \ \zeta^2/ (\beta \varrho_o)^2\rho_f$.
    
    In the case of active flow-driven communication, relevant for social insects, we assume that the agents operate in the interesting neighborhood of the threshold pheromone concentration $c \sim c^*$, we therefore replace $\Theta(\delta c) \rightarrow \R_\varrho \delta c$.  In the limit of rapid equilibration of the pheromone concentration and a rapid relaxation to steady velocity relative to agent recruitment rates and pheromone decay rates relevant for social insects, we assume that $\partial_t c \approx 0$ and $\partial_t \u_f \approx 0$, so that Eqs.~\ref{eq:rhob}-\ref{eq:incomp} in scaled units reduce to
    \begin{align}
    	\partial_t \varrho + \V_\varrho \nabla_{||} \cdot (\varrho \grad_{||} c) =& \ \D_\varrho \nabla_{||}^2 \varrho +
    	( 1 - {\varrho}/{\varrho_\max}) [ (1-\varrho) \nonumber \\
    	& \ + \R_\varrho  \delta c ], \label{eq:mNDrhob} \\
    	 \D_c \nabla^2 c + \V_c \grad \cdot (\u_f c) =& \ \R_c \varrho - c, \label{eq:mNDcp} \\
    	\D_u (\nabla^2 \u_f - \u_f) =& \ \grad p \label{eq:mNDuf}\\
    	\nabla \cdot \u_f =& \ 0. \label{eq:mNDcont}
    \end{align}
    When the diffusion of pheromone dominates over advection i.e. $\D_c \sim \R_c \gg 1, \V_c$ and in the absence of recruitment and relaxation ($\alpha_p, k_b = 0$) or when the agent density has reached the maximum value ($\varrho = \varrho_{\max}$), Eqs.~\ref{eq:mNDrhob}-\ref{eq:mNDcp} reduce to the PKS equations (see SI sec.~IV B for details). In this article, however, we are interested in the limit relevant for social insects where the dynamics of the agents is governed primarily by chemotaxis and recruitment, i.e. $\D_\varrho \ll \V_\varrho \sim \R_\varrho$ and that the pheromone rapidly equilibrates due to advection, i.e. $\D_c \ll \V_c \sim \R_c$. From now on, we will therefore ignore the effects of agent and pheromone diffusivity.
    
    We explore the different limits of the phase-space defined by the pheromone dynamics (governed by $\V_c, \R_c$) and by agent dynamics on the other (governed by $\V_\varrho, \R_\varrho$) to understand the different modes of collective patterning determined by our model. {There are four different regimes characterized by relative magnitudes of $\R_\varrho, \V_\varrho$ and $\R_c, \V_c$ (see Fig.~\ref{fig:phasediag}$(c)$) which we address in a 1-dimensional setting before turning briefly to the 2-dimensional case (see SI sec.~IV for details).}\\
    
    \begin{figure}
    	\centering
    	\includegraphics[width=0.46\textwidth]{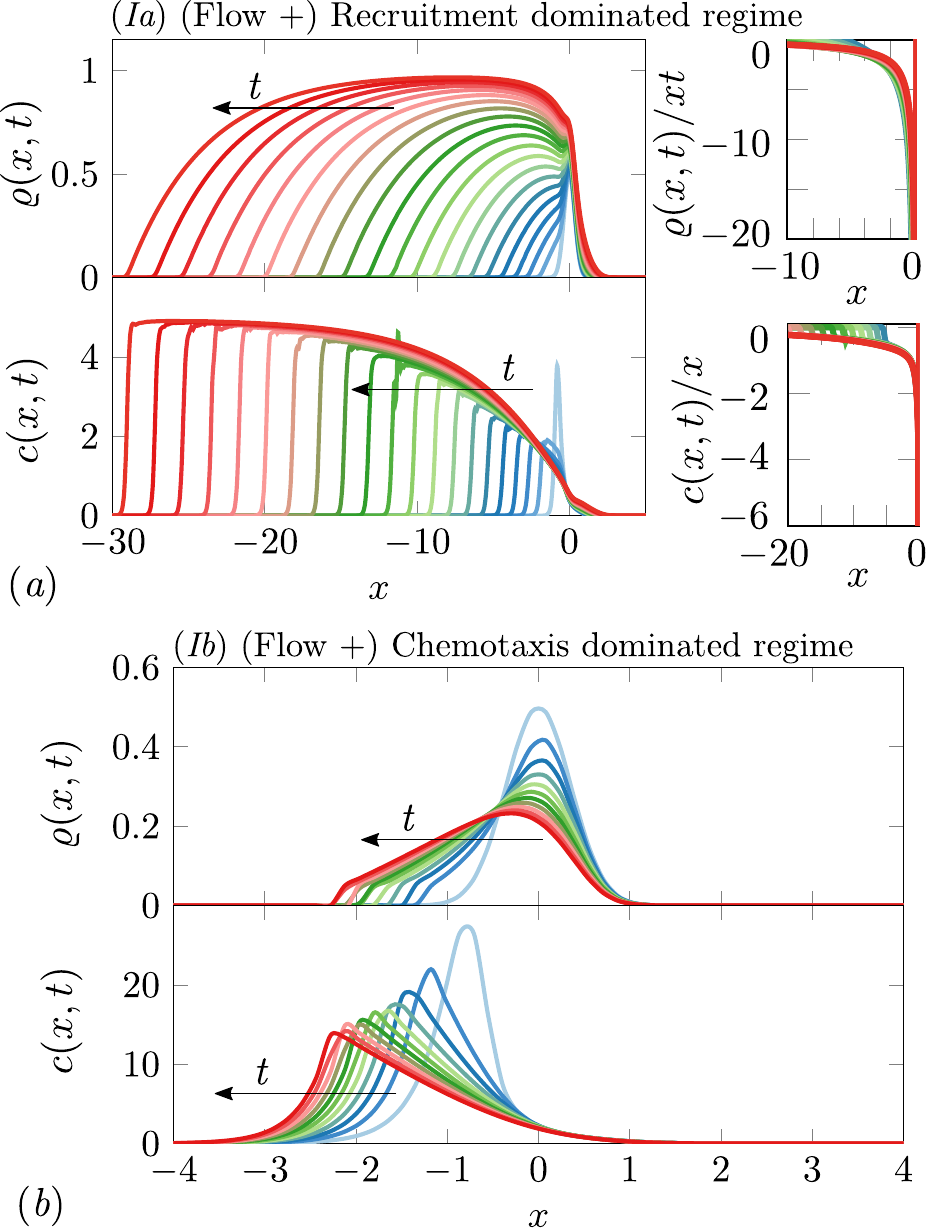}
    	\caption{Dynamics of agent density, $\varrho (x,t)$ and pheromone concentration, $c (x,t)$ in the $(a)$ flow (and recruitment) dominated regime (\textit{Ia} in the text) where agents polarize in response to autogenic flow and recruitment, leading to an increase in the agent density (for $\V_c = 0.5, \R_c = 0.5, \V_\varrho = 0.0, \R_\varrho = 300.0$). The plots on the right show the self-similar evolution of agent density and the pheromone concentration, consistent with a the collapse of the different curves when scaled by $xt, x$ before saturation (see SI sec.~IV A for details). $(b)$  Flow (and chemotaxis) dominated regime phase (\textit{Ib} in the text) where the agents migrate upstream in the direction of flow through (for $\V_c = 5, \R_c = 0.9, \V_\varrho = 280.0, \R_\varrho = 0.0$). The results were obtained by numerically solving Eqs.~\ref{eq:rhoRed}, \ref{eq:cpRed} (see SI sec.~V for details).  }
    	\label{fig:1Ddyn}
    \end{figure}
    
    \noindent \textit{(I) Flow dominated regimes} ($\V_c, \R_c \gg 1$) : {In the flow-dominated regime, we explore the collective behavior of the agents as we vary $\R_\varrho/\V_\varrho$ - see  Fig.~\ref{fig:phasediag}$(c)$}. Assuming that the colony is uniformly active and generating flow everywhere, i.e. $\varrho = 1$, the velocity field is given by the solution of Eqs.~\ref{eq:mNDuf},~\ref{eq:mNDcont} subject to the shear boundary condition at the wall: $\partial_z (\u_f\cdot \hat{\x}_{||}) (z=0) = 1$, which yields a polarized flow field in the in-plane direction $u_{||} (z) \equiv \u_f \cdot \hat{\x}_{||} = - \exp(-z)$. Then the pheromone concentration (obtained by substituting the above result in Eq.~\ref{eq:mNDcp}) yields $c(x)~=~\R_c [1 - \exp{(x/\V_c)}]$. This inhomogeneous pheromone concentration profile driven by a homogeneous agent density $\varrho=1$ causes the agents to migrate when the chemotactic response $\V_\varrho > 0$ (see from Eq.~\ref{eq:mNDrhob}), indicating the unstable nature of this density profile.
    
    Substituting the exponential flow profile into Eq.~\ref{eq:mNDcp} for the pheromone dynamics (see SI sec.~III for details) in the limit of advection-dominated pheromone transport leads to the shear becoming proportional to agent density so that $\partial_{z} u_{||} \sim 1$, and Eqs.~\ref{eq:mNDrhob}-\ref{eq:mNDuf} simplify to the coupled system for the agent density and the pheromone concentration,
    \begin{align}
		\partial_t \varrho + \V_\varrho \partial_x (\varrho \partial_x c) =& \ (1-\varrho/\varrho_\max)[(1 - \varrho) + \R_\varrho c], \label{eq:rhoRed}\\
		-\V_c \partial_x (\varrho c) =& \ \R_c \varrho - c. \label{eq:cpRed}
    \end{align}
    To explore the behavior of the collective agent dynamics in the different limits of phase-space we solve Eqs.~\ref{eq:rhoRed},~\ref{eq:cpRed} (see SI sec.~V for details).
    
    \noindent \textit{(Ia) Recruitment dominated regime} ($\R_\varrho \gg 1, \V_\varrho \ll 1$):  In Fig.~\ref{fig:1Ddyn}$(a)$ we see that an initially localized gaussian profile of the agent density $\varrho(x,0)$ centered around $x=0$ grows and spreads along the direction of pheromone advection due to flow-driven agent recruitment.  Eventually, the agent density saturates to maximal packing density $\varrho=\varrho_{\max}$. Simultaneously, the pheromone concentration $c(x,t)$ increases with increasing recruitment and is also advected by flow (see Fig.~\ref{fig:1Ddyn}$(a)$). To understand the scaling behavior of the spatio-temporal profiles of the agent and pheromone density in the growth phase, we note that when the pheromone is produced by agents and transported by the active flow with decay being insignificant,  Eq.~\ref{eq:cpRed} reduces to $-\V_c \partial_x (\varrho c) \sim \R_c \varrho$. Furthermore, when the effects of chemotaxis are small (i.e. $\V_\varrho \rightarrow 0$), the agent density scales as $\varrho \sim x t$ while the concentration of the pheromone scales as $c \sim x$ (see SI sec.~IV A for a detailed analysis). Normalizing $\varrho(x,t)$ and $c(x,t)$ using this scaling causes the curves to collapse on top of each other (see rightmost figures in Fig.~\ref{fig:1Ddyn}$(a)$).
    
    \noindent \textit{(Ib) Chemotaxis dominated regime} ($\V_\varrho \gg 1, \R_\varrho \ll 1$):   In Fig.~\ref{fig:1Ddyn}$(b)$ we see that an initially localized agent density collectively migrates up the pheromone gradient. Simultaneously the concentration of pheromone decreases in magnitude and migrates with the agents. To understand these observations we note that in the flow-dominated regime,  Eq.~\ref{eq:cpRed} implies that $-\V_c \partial_x (\varrho c) \sim \R_c \varrho$, which when substituted into Eq.~\ref{eq:rhoRed} yields: $\partial_t \varrho - (\V_\varrho \R_c /\V_c) \partial_x \varrho = \V_\varrho \partial_x(c\partial_x \varrho)$. We see that agents migrate with a speed $(\V_\varrho \R_c /\V_c)$ and diffuse inhomogeneously due to the pheromone concentration that acts as a spatially varying diffusivity. Here the emergence of collectively polarized  migration even in the absence of recruitment ($\V_\varrho \gg 1, \R_\varrho \rightarrow 0$) is due to autonomous fluid flow coupled with pheromone production.\\
    
    \noindent \textit{(II) Pheromone decay dominated regimes} ($\V_c \ll \R_c \sim 1$): 
    In the pheromone dominated regime, we explore the collective behavior of the agents as we vary $\R_c/\V_c$ - see Fig.~\ref{fig:phasediag}$(c)$. In the limit of a homogeneous agent density $\varrho = 1$,  when the effects of decay dominate ($\R_c \gtrsim 1, \V_c \rightarrow 0$), Eq.~\ref{eq:cpRed} yields $c^\st = \R_c$. Perturbing about this steady state ($\varrho^\st = 1, c^\st = \R_c$) using the ansatz $\{ \varrho(x,t) - \varrho^\st, c(x,t) - c^\st \} = \{ \rhoh (q), \ch (q) \} \exp(iqx+\Omega t)$ where $\Omega(q)$ is the frequency of oscillation and $q$ the scalar wavenumber, $\rhoh (q), \ch(q)$ are the complex amplitudes of agent density and pheromone concentration and substituting into Eqs.~\ref{eq:mNDrhob},~\ref{eq:mNDcp}  yields the dispersion relation (derived in SI sec.~II),
    \begin{align}
        \Omega(q) =& \ \frac{1}{2}\bigg\{- (1 + \lambda) \pm
       [(\lambda - 1)^2 + 4 \lambda \R_c( \R_\varrho + 4 q^2 \V_\varrho)]^{1/2} \bigg\},\label{eq:disp}
    \end{align}
    where we have $\lambda = (k_-/k_b)$ and we use $k_b^{-1}$ as the time-scale to non-dimensionalize the frequency. When $\Re(\Omega(q)) > 0$ the agent density becomes unstable; for long wavelengths ($q \rightarrow 0$) Eq.~\ref{eq:disp} becomes $\Omega(q) \approx [-(1 + \lambda) + \sqrt{(\lambda-1)^2 + 4 \lambda \R_c \R_\varrho}]/2$ and we see that the dynamics is governed by competition between recruitment sensitivity (from $\R_\varrho$) and  pheromone decay (from $\R_c$). When $\R_\varrho \gg 1$ corresponding to large recruitment sensitivity this triggers long wavelength perhaps suggestive of such a mechanism at play when the colony in under threat from an intruder. For smaller wavelengths ($q \gg 1$) the instability is due to competition between pheromone decay (contributions from $\R_c$) and chemotaxis (from $\V_\varrho$), as $\Omega(q) \approx  2 \sqrt{\lambda \R_c \V_\varrho} q$. This linearly growing dependence of the unstable frequency on wavenumber is saturated by the effects of diffusion at large wavenumbers as $\Omega(q) \approx -\D_c \lambda q^2$.
    
    \begin{figure}
    	\centering		\includegraphics[width=0.45\textwidth]{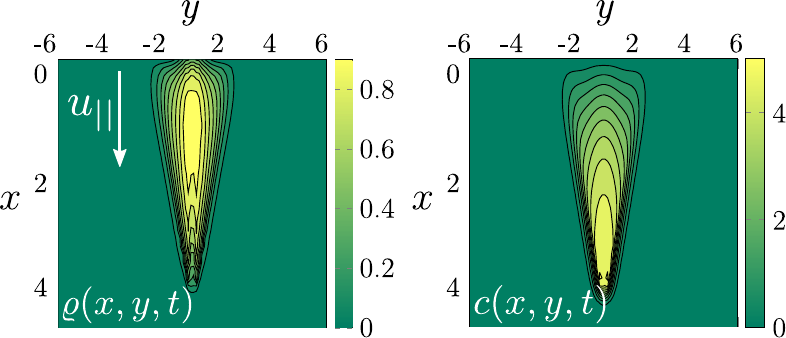}
    	\caption{Polarized recruitment of active agents and corresponding pheromone concentration in two-dimensions obtained by solving Eqs.~\ref{eq:mNDrhob},~\ref{eq:cpRed} and plotted at $t = 1.5$. Agent density along $x=0$ is set to $\varrho(0,y,t) = \exp(-y^2/2w^2)$ with $w = 2$ and these agents generate flow $u_{||}$ along positive $x$-axis that results in recruitment of other agents. The triangular shape is a consequence of chemotaxis driven by gradients in pheromone along $y$-axis. We choose Neumann boundary condition for $\varrho(x, y, t), c(x, y, t)$ along other boundaries. The non-dimensional numbers are chosen to be $\V_c = 0.5, \R_c = 0.5, \V_\varrho = 25.0, \R_\varrho = 300.0$ (see SI sec.~V for further details).}
    	\label{fig:2Ddyn}
    \end{figure}
    To understand the nonlinear evolution of the agents in the pheromone-decay dominated regime, the concentration of the pheromone is proportional to the density everywhere, $c \approx \R_c \varrho$. Substituting this into Eq.~\ref{eq:cpRed} we get,
    \begin{align}
    	\partial_t \varrho + (\V_\varrho \R_c/2) \partial^2_{y} \varrho^2 =& \ \R_\varrho \R_c \varrho + (1 - \varrho),\label{eq:nonLin}
    \end{align}
    The agent density can exhibit two distinct behaviors, determined by the relative magnitude of chemotaxis and recruitment.
    
    \noindent \textit{(IIa) Chemotaxis dominated regime} ($\V_\varrho \gg 1, \R_\varrho \ll 1$): 
     In this limit, Eq.~\ref{eq:nonLin} reduces to a nonlinear heat equation with a negative diffusivity. As the agents sense the pheromone generated by their immediate neighbors (as $\V_\varrho \gg 1$) they migrate towards each other which results in a pile-up of the density (see SI sec.~IV A for details of the similarity solution associated with aggregation). This aggregation behavior is qualitatively similar to what is seen in the PKS model, however the dynamics follows a different scaling law.
    
    \noindent \textit{(IIb) Recruitment dominated regime} ($\R_\varrho \gg 1, \V_\varrho \ll 1$): In this limit, Eq.~\ref{eq:nonLin} reduces to a linear homogeneous equation and the agent density grows exponentially as $\varrho \sim \exp(\R_\varrho \R_c t)$. This leads to homogeneous growth (see SI Fig.~S2$(d)$)  unlike the flow-dominated scenario where advection of the pheromone leads to polarized recruitment and anisotropic patterning.\\
    
    \noindent \textit{Analysis in 2D:} Our 1D setting and analysis translates naturally to 2D scenarios. When recruitment is driven by fluid flow ($\V_c \sim \R_c$), we expect accumulation along the transverse direction ($x$-axis) when $\R_\varrho \gg 1$ as shown in Fig.~\ref{fig:2Ddyn}, with a concomitant  narrowing of the accumulated agent density along $y$-axis seen in Fig.~\ref{fig:2Ddyn}. This results in the evolution of the density and the pheromone concentration towards a triangular shape seen in recent experiments with bees~\cite{nguyen2021flow} (shown in Fig.~\ref{fig:1Ddyn}$(a)$ inset).
    
    Our study of the spatio-temporal organization of a collective of agents that actively generate and advect pheromones using autonomous flows to recruit additional agents generalizes the classic Patlak-Keller-Segel diffusive aggregation framework  and shows the emergence of  spatio-temporal patterns of agents that collectively communicate, migrate and aggregate. We delineate this in a phase-space defined by the ratio of recruitment strength to chemotaxis i.e. $\R_\varrho/\V_\varrho$ and the ratio of shear strength to pheromone decay strength i.e. $\V_c/\R_c$. For large shear strength and recruitment strength ($\V_c/\R_c, \R_\varrho/\V_\varrho \gg 1$) the agent density exhibits successful polarized recruitment driven by the autogenic fluid flow. On the other hand, when the chemotaxis strength dominates ($\V_c/\R_c, \V_\varrho/\R_\varrho \gg 1$) the agents are able to collectively migrate along the direction of the fluid flow. When the pheromone decay effects dominate ($\R_c/\V_c, \V_\varrho/\R_\varrho \gg 1$) agents collectively aggregate which eventually results in the formation of shocks. Lastly, when the chemotactic velocity is weak but the recruitment strength is large ($\R_c/\V_c, \R_\varrho/\V_\varrho \gg 1$) the agents can perform isotropic recruitment. Generalizing our framework to include the orientation dynamics of agents while accounting for the role of noise in sensing and actuation are natural next steps in characterizing decision making and collective response in active, embodied agents that use rheomergy.
    
    \acknowledgments We thank Orit Peleg for Fig.~\ref{fig:phasediag}$(a)$, and the NSF Simons Center for Mathematical and Statistical Analysis of Biology, the Simons Foundation and the Seydoux Fund for partial support.

    \bibliographystyle{apsrev4-1}
    \bibliography{biblio}
\end{document}